\documentclass[aps,prl,twocolumn,groupedaddress,showpacs,floatfix,superscriptaddress,longbibliography]{revtex4-2}
\usepackage{epsfig}
\usepackage{amsmath,amssymb}
\usepackage{graphicx}
\usepackage[dvipsnames,usenames]{xcolor}
\usepackage[normalem]{ulem}
\usepackage{soul}  
\usepackage[hyperindex,breaklinks,colorlinks=true, citecolor = blue]{hyperref}
\emergencystretch=\maxdimen
\newcommand{\im}{{\rm i}}

\begin{document}

\title{Attractive Su-Schrieffer-Heeger-Hubbard Model on a Square Lattice Away from Half-Filling}

\author{Bo Xing}
\email{bo\_xing@mymail.sutd.edu.sg}
\affiliation{Science, Mathematics and Technology Cluster, Singapore University of Technology and Design, 8 Somapah Road, 487372 Singapore}

\author{Chunhan Feng}
\affiliation{Center for Computational Quantum Physics, Flatiron Institute, 162 5th Avenue, New York, New York, USA}
\affiliation{Department of Physics, University of California, Davis, CA 95616, USA}

\author{Richard Scalettar}
\affiliation{Department of Physics, University of California, Davis, CA 95616, USA}

\author{G. George Batrouni}
\affiliation{Centre for Quantum Technologies, National University of Singapore; 2 Science Drive 3 Singapore 117542}
\affiliation{Department of Physics, National University of Singapore, 2 Science Drive 3, 117542 Singapore} 
\affiliation{Universit\'e C\^ote d'Azur, CNRS, Institut de Physique de Nice (INPHYNI), 06000 Nice, France}

\author{Dario Poletti}
\affiliation{Science, Mathematics and Technology Cluster, Singapore University of Technology and Design, 8 Somapah Road, 487372 Singapore}
\affiliation{EPD Pillar, Singapore University of Technology and Design, 8 Somapah Road, 487372 Singapore}
\affiliation{Centre for Quantum Technologies, National University of Singapore; 2 Science Drive 3 Singapore 117542}

\date{\today}

\begin{abstract} 
The Su-Schrieffer-Heeger (SSH) model, with bond phonons modulating electron tunneling, is a paradigmatic electron-phonon model that hosts an antiferromagnetic order to bond order transition at half-filling.
In the presence of repulsive Hubbard interaction, the antiferromagnetic phase is enhanced, but the phase transition remains first-order. 
Here we explore the physics of the SSH model with {\it attractive} Hubbard interaction, which hosts an interesting interplay among charge order, $s$-wave pairing, and bond order.
Using the numerically exact determinant quantum Monte Carlo method, we show that both charge order, present at weak electron-phonon coupling, and bond order, at large coupling, give way to $s$-wave pairing when the system is doped. 
Furthermore, we demonstrate that the SSH electron-phonon interaction competes with the attractive Hubbard interaction and reduces the $s$-wave pairing correlation. 
\end{abstract}

\maketitle

\noindent
\underline{{\it Introduction:}}
Electron-electron and electron-phonon correlations are two fundamental interactions in condensed matter quantum many-body systems.
The basic qualitative features of the electron-electron interaction are famously studied in the Hubbard model~\cite{Hubbard1963, gutzwiller1963, kanamori1963, hubbard1964, hubbard1964a}, which, depending on interaction's magnitude and doping, can exhibit metallic phases, long-range spin/charge patterns, and superconducting pairing~\cite{tasaki1998hubbard,georges96, LeeWen2006,qingull22,arovas2022hubbard}.
Similarly, the phenomena induced by the electron-phonon interaction are often studied in the Holstein~\cite{holstein59} and Su-Schrieffer-Heeger (SSH)~\cite{su79} models.
The former considers site phonon vibrations that influence the on-site chemical potential and can result in charge density wave (CDW) and $s$-wave pairing~\cite{Noack1991, Vekic1992, Zhang2019, CChen2019, Feng2020, Weber2018, Cohen-Stead2019, ZXLi2019, Feng2020i, Nosarzewski2021, Xiao2021, Bradley2021, dee2020relative, Nosarzewski2021,zhang2022,kvande2023}.
The latter considers bond phonon vibrations that modulate nearest-neighbor tunneling and give rise to antiferromagnetic order and bond order wave (BOW)~\cite{fradkin83, fradkinhirsch1983, hirsch83, zheng88, mckenzie96, sengupta03, barford06, pearsonbursill11, bakrim15, weber15, Weber2020, XingBatrouni2021, CaiYao2021, GoetzAssaad2021}.

The interplay between the electron-electron and the electron-phonon interactions has attracted a lot of attention.  
Extensive studies on the repulsive Hubbard-Holstein model at half-filling have shown that magnetic and charge order dominates in the strong electron-electron and electron-phonon interaction regimes respectively~\cite{bergerlindon1995, freericksjarrell1995, honerkamplee07, nowadnickdevereaux12, wangwang15}. 
Where the electron-electron and electron-phonon interactions are comparable, numerical simulations reveal a metal or superconducting pairing phase~\cite{JohnstonDeveraux2013, ohgoeimada17, karakuzubecca17, WangDemler2020}.  
Away from half-filling, Monte Carlo simulations are severely undermined by the sign problem~\cite{loh90, gubernatiszhang94,troyer2005computational,mondaini2022quantum,tarat2022deconvolving}, and numerical evidence predicts either superconducting pairing or stripe order~\cite{ohgoeimada17, KarakuzuJohnston2022}.

The combined effects of SSH phonons with Hubbard interactions have only been more recently
investigated. The repulsive SSH-Hubbard (SSHH) model on a square lattice features a direct first-order transition between antiferromagnetic order and BOW at half-filling~\cite{fengbatrouni22, cai_2021}. 
The repulsive SSHH model also suffers from the sign problem away from half-filling.
Nonetheless, a density matrix renormalization group study on narrow 2D cylinders and a functional renormalization group study of square lattices have shown evidence of superconducting pairings around $1/8$ doping~\cite{WangYao2022, YangWang2022}.

In the above studies, the on-site electron-electron interaction was taken to be repulsive.
However, a similar, and potentially rich, phenomenology is  expected at the interplay between electron-phonon interaction and {\it attractive} on-site electron-electron interactions.
In particular, the attractive Hubbard model exhibits a competition between charge order and s-wave pairing~\cite{hirsch1985, scalettardagotto1989, hogiamarchi09, chankohl20}, with bond order due to SSH phonons.
Importantly, the attractive SSHH model does not suffer from the sign problem.
This makes it a rare electron-electron and electron-phonon model that can be studied exactly at all fillings and low temperatures.

In this paper, we present a study of the single-orbital square lattice attractive SSHH model with periodic boundary conditions.
We consider the full phonon dynamics and focus on the phase transitions away from half-filling.
Due to the absence of the sign problem, we can simulate systems of up to $12 \times 12$ spatial sites at low temperatures using the determinant quantum Monte Carlo (DQMC) method~\cite{blankenbecler81, scalettar89, noack91, batrouni19a, GubernatisWerner2016}.
First, we show the effects of attractive interaction in the half-filled BOW ground state.
In addition, we analyze the effect of temperature and phonon frequency on the ground state phase transitions. 
At half-filling, the system presents either a charge density wave or bond order depending on the magnitude of the electron-phonon interactions. Larger attractive interactions reduce the bond order. Away from half-filling the system abruptly goes into an $s$-wave pairing ordered phase independent of the magnitude of the electron-phonon interaction. 

\vskip0.05in \noindent
\underline{{\it Model and method:}}
The attractive SSHH Hamiltonian is,
\begin{align}\label{eq:SSHHneg_ham}
    \mathcal{\hat H} =& - t \sum_{\langle i, j \rangle, \sigma} \left( 1 - \lambda \hat{X}_{ij} \right) \left( \hat{c}^{\dagger}_{i \sigma} \hat{c}^{\phantom\dagger}_{j \sigma} + {\mathrm{h.c.}} \right) - \mu \sum_{i, \sigma} \hat{n}_{i \sigma} \nonumber \\
                 &+ U \sum_{i} \left(\hat{n}_{i\uparrow}-\frac{1}{2}\right) \left(\hat{n}_{i\downarrow}-\frac{1}{2}\right) \nonumber \\
                 & + \sum_{\langle i, j\rangle} \left(\frac{1}{2M} \hat{P}_{ij}^2 + \frac{M}{2} \omega_{0}^2 \hat{X}_{ij}^2 \right)
,\end{align}
where $\hat{c}_{i \sigma}$ ($\hat{c}_{i \sigma}^{\dagger}$) destroys (creates) an electron of spin $\sigma=\uparrow,\downarrow$ on site $i$, $U$ is the strength of electron-electron interaction, $\hat{n}_{i\sigma} = \hat c^{\dagger}_{i\sigma}\hat c^{}_{i\sigma}$ is the number operator, $M$ is the phonon mass and $\omega_{0}$ the phonon frequency. $\hat X_{ij}$ and $\hat P_{ij}$ are the canonically conjugate phonon displacement and momentum.
The dimensionless electron-phonon coupling strength is $g=\lambda/\sqrt{2M\omega_{0}/\hbar}$. 
When $U < 0$, the Hubbard interaction is attractive.
The chemical potential, $\mu$, controls the density of the system.
The system is at half-filling, $\langle n \rangle = \sum_i\langle \hat n_{i \uparrow} + \hat n_{i \downarrow} \rangle/N = 1$, for $\mu = 0$.
In an ordered phase, a gap can open such that the system leaves half-filling only beyond a finite critical $\mu_{c}$.
The value of $\mu_{c}$ depends on the other system parameters.
Here, we work in units where $\hbar=M=t=k_{B}=1$ ($k_{B}$ is the Boltzmann constant).

\begin{figure}[t]
    \centering
    \includegraphics[width=\linewidth]{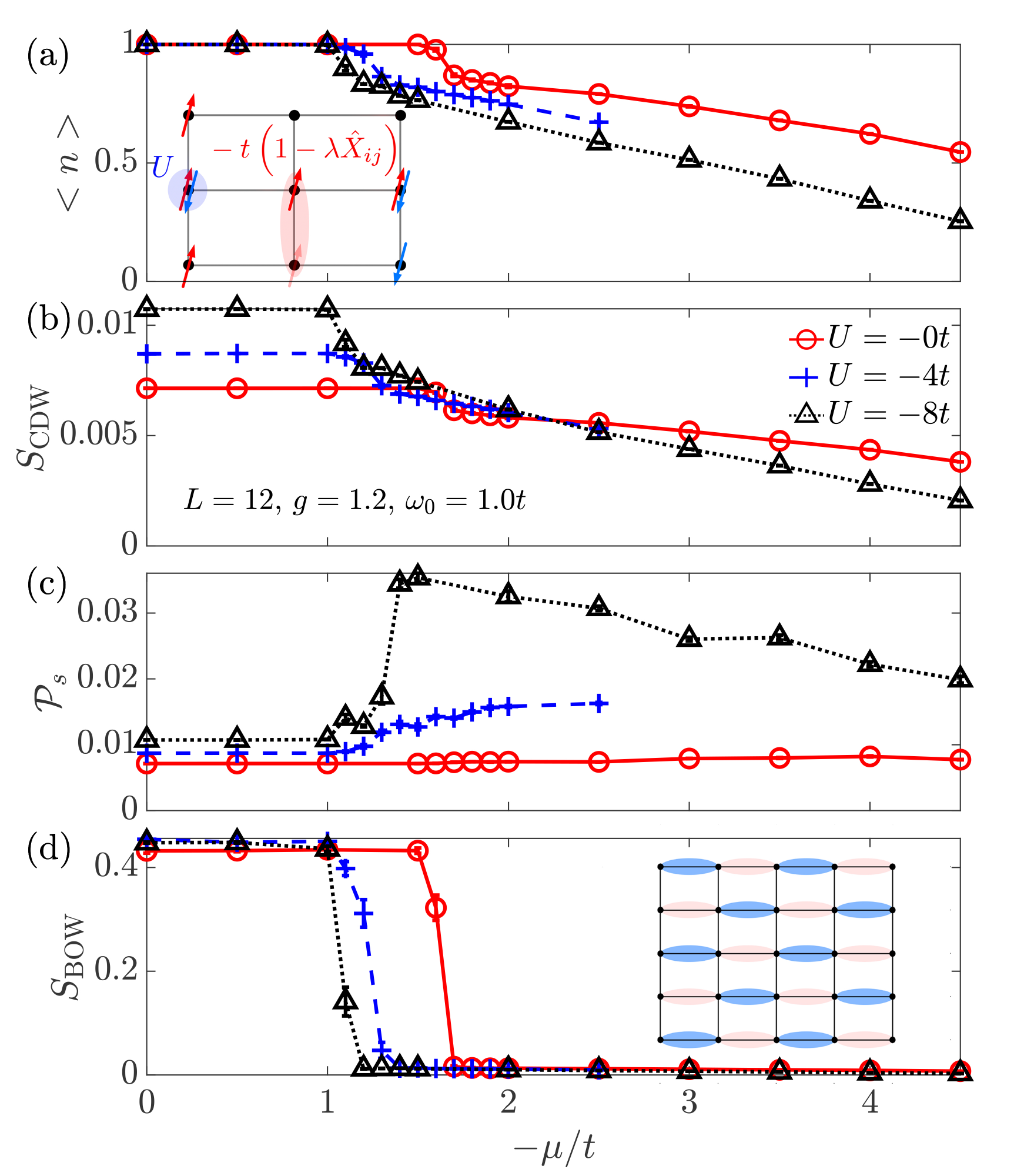} 
    \caption{
        (a) Average density $\langle n \rangle$, (b) CDW structure factor $S_{\mathrm{CDW}}$, (c) $s$-wave pairing correlation $\mathcal{P}_{s}$ and (d) BOW structure factor $S_{\mathrm{BOW}}$ as functions of chemical potential $\mu$.
        Inset of panel (a): graphical representation of the electron-electron and electron-phonon interactions.
        Inset of panel (d): graphical representation of the $q = \left( \pi, \pi \right)$ BOW. Different bond colors emphasize the alternation in bond lengths. 
        In all panels $L = 12$, electron-phonon coupling $g = 1.2$ and phonon frequency $\omega_{0} = 1.0t$.
    }
    \label{fig:fixg_varyU_b16}
\end{figure}

The attractive SSHH model does not suffer from the sign problem because both the SSH phonons, the chemical potential, and the Hubbard-Stratonovich~\cite{Stratonovich1957, negeleorland1988, fulde1995, Enz1992, fetterwalecka03} field couple to the spin-up and spin-down electrons identically.  Consequently, the determinants that arise from integrating the fermionic degrees of freedom are identical for the two spin species, and the configurational weight is a perfect square.
We work with inverse temperature $\beta = L_{\tau}\Delta\tau = 16t,\ 21t,\ 24t$, where $L_{\tau} = 300$ is the number of imaginary time steps.
Even for the largest value of $\Delta \tau = 0.08t$, the estimate of the Trotter error is of the same order as the statistical error bars in the structure factors.
In addition, to elucidate finite size effects, we compare three $L \times L$ systems with $L= 8, 10, 12$.

To characterize the underlying ground state phases in different parameter regimes, we calculate the CDW correlations and structure factors: 
$C_{\mathrm{CDW}} \left( i-j \right) = \langle \left( \hat n_{i,\uparrow} + \hat n_{i,\downarrow} \right) \left( \hat n_{j,\uparrow} + \hat n_{j,\downarrow} \right) \rangle$ 
and their Fourier transform  
$S_{\mathrm{CDW}} \left( q \right) = \frac{1}{N}\sum_{i,j}e^{\im q \left( i-j \right)}  C_{\mathrm{CDW}}\left( i-j \right)$; 
and the BOW correlations and structure factors: 
$C_{\mathrm{BOW}}  (i-j) =
 \langle  \hat c_{i,\sigma}^{\dagger}\hat c_{i+\hat{x}(\hat{y}),\sigma}^{}   
\hat c_{j,\sigma}^{\dagger} \hat c_{j+\hat{x}(\hat{y}),\sigma}^{} \rangle$ 
and its Fourier transform 
$S_{\mathrm{BOW}}(q) = \frac{1}{N}\sum_{i,j}e^{\im q \left( i-j \right)} C_{\mathrm{BOW}}\left( i-j \right)$ respectively.
The various wave pairing correlations are characterized by $\mathcal{P}(i-j) = \langle \hat \Delta(i+j)^{}\hat \Delta^{\dagger}(i) \rangle$~\cite{whitescalettar89}.
For the case of $s$-wave pairing $\hat \Delta^{\dagger}(i) = \hat c^{\dagger}_{i,\uparrow} \hat c^{\dagger}_{i,\downarrow}$. 
Average density $\langle  n \rangle$ is also useful in giving insight into the properties of the phases. 

\vskip0.05in \noindent
\underline{{\it Results:}}
Regardless of the sign of $U$, on a bipartite lattice, performing a (staggered) particle-hole transformation on both electron species
changes the sign of the chemical potential while leaving the Hamiltonian and the spin, charge, and pair correlations invariant.
As a consequence, the phase diagram is symmetrical under reflection about the point 
$\langle n \rangle = 1$.
Furthermore, on a bipartite lattice precisely at $\langle  n \rangle = 1$ $(\mu=0)$, 
the repulsive SSHH model can be transformed into the attractive SSHH model by 
performing a {\it partial} particle-hole transformation on one of the two electron species.
Therefore, the phase diagram of the attractive model can be inferred from that of
the repulsive model at $\langle n \rangle = 1$.
Importantly, via the partial particle-hole transformation the magnetic order in the repulsive SSHH model is transformed into $s$-wave pairing correlations and charge order in the attractive SSHH model at half-filling.

Away from half-filling, the repulsive SSHH model no longer directly maps to the attractive SSHH model.
We first investigate the effects of the electron-electron coupling, $U$, in a situation of large electron-phonon coupling, $g=1.2$, 
where the system is firmly in the SSH BOW phase at half-filling. 
We will then consider the effect of the electron-phonon coupling, $g$, on the degenerate CDW/$s$-wave pairing phase which arises when $U$ is the dominant energy scale.

In Fig.~\ref{fig:fixg_varyU_b16}, we fix the electron-phonon coupling $g = 1.2$, phonon frequency
$\omega_{0} = 1.0 \, t$, vary the chemical potential $\mu$, and investigate the effects of the electron-electron coupling $U$.
At $\mu = 0$, the ground state is a gapped $q = \left(\pi, \pi\right)$ BOW in either the $x$ or $y$ direction for all $U/t = 0,\ -4,\ -8$ (Fig.~\ref{fig:fixg_varyU_b16}(a)).
When $|\mu|$ is lower than the critical $|\mu_{c}|$, the average density is pinned at $\langle n \rangle = 1$ and the system stays in a gapped $\left(\pi, \pi\right)$ BOW phase.
The CDW structure factor, $S_{\mathrm{CDW}}$, and $s$-wave pairing correlation, $\mathcal{P}_{s}$, are negligible when compared to the BOW structure factor, $S_{\mathrm{BOW}}$.
At finite $U$ and $\mu = \mu_{c}$, there is a noticeable increase of s-wave pairing $\mathcal{P}_{s}$ (Fig.~\ref{fig:fixg_varyU_b16}(c))~\footnote{Other electron-electron pairings, such as the $d$-wave pairing, are much lower than the $s$-wave pairing (not shown).}. 
In addition, $\langle n \rangle$ starts to decrease and $S_{\mathrm{BOW}}$ falls discontinuously within a narrow range of $\Delta \mu$ in Fig.~\ref{fig:fixg_varyU_b16}(d).
This suggests a first-order transition from the BOW phase to $s$-wave pairing in the attractive SSHH model.
As $U$ increases in magnitude, $\mu_{c}$ shifts towards $\mu = 0$ and $s$-wave pairing correlations increase.
Interestingly, the BOW structure factor is also slightly enhanced in the presence of $U$ before the transition.
There is a qualitative change in $S_{\mathrm{CDW}}$ at $\mu_{c}$ as shown in Fig.~\ref{fig:fixg_varyU_b16}(b).
For this (large) value of electron-phonon coupling, $g=1.2$, there is no CDW order at any filling
including $\langle n \rangle =1$.

\begin{figure}[t]
    \centering
    \includegraphics[width=\linewidth]{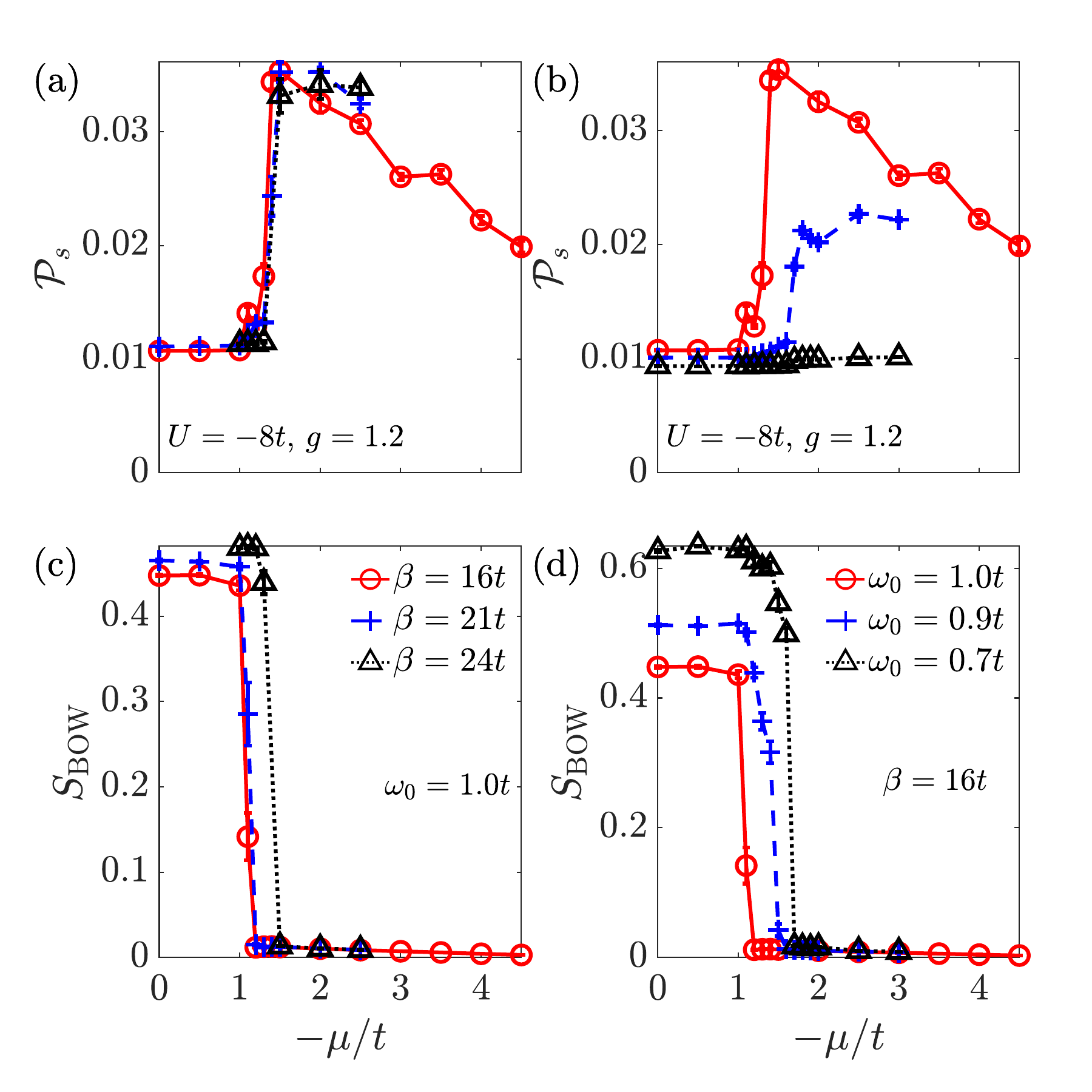} 
    \caption{
        (a,b) $s$-wave pairing correlation $\mathcal{P}_{s}$ and (c,d) BOW structure factor $S_{\mathrm{BPW}}$ against chemical potential $\mu$.
        For all results, $L = 12$ and electron-phonon coupling $g = 1.2$.
        (a,c) $\beta$ is varied and $\omega_{0}$ is fixed.
        (b,d) $\omega_{0}$ is varied and $\beta = 16$.
    }
    \label{fig:fixUg_varyo_varyb}
\end{figure}

Next, we show that the inverse temperature $\beta$ we use is large enough to capture the ground state properties and study the effect of the phonon frequency $\omega_{0}$ on the phase transition.
In Fig.~\ref{fig:fixUg_varyo_varyb}, we fix $U = -8\,t$, $g = 1.2$ and show structure factor data for different $\beta$ (a,c) and $\omega_{0}$ (b,d). 
For panels (a,c), $\omega_0 = 1.0\,t$ and $\beta$ is varied.
There are no qualitative and only minor quantitative changes in the structure factors and $\mu_{c}$ as $\beta \, t$ increases from $16$ to $24$.
The transition between BOW and $s$-wave pairing remains first-order and $\mu_{c}$ increases very slightly.
For panels (b,d), $\beta = 16/t$ and $\omega_{0}$ is varied.
At $\omega_{0} = 0.7$, we see a similar rapid decrease in $S_{\mathrm{BOW}}$.
The increase in $\mathcal{P}_{s}$ is much less apparent.
However, we cannot rule out the existence of a small nonzero pairing order at small $\omega_0$.
Numerical evidence of $s$-wave pairing has always been difficult in electron-phonon models, especially at lower $\omega_{0}$.
In the Holstein model, $s$-wave pairing is known to increase with $\omega_{0}$~\cite{Nosarzewski2021,hohenadlerbatrouni19,ohgoeimada17}.
At low $\omega_{0}$, the order parameter is likely to be small, if it exists.
The situation is similar in the attractive SSHH model.
At higher $\omega_{0}$, the lattice responds more quickly to electronic hopping, and the electron-phonon fluctuation is suppressed.
As a result, $\mathcal{P}_{s}$ is higher.
The critical $\mu_{c}$ needed to leave half-filling decreases and so does $S_{\mathrm{BOW}}$ within the BOW phase.
For all $\omega_{0}$ that support BOW at half-filling, the BOW phase transitions directly to $s$-wave pairing.

\begin{figure}[t]
    \centering
    \includegraphics[width=\linewidth]{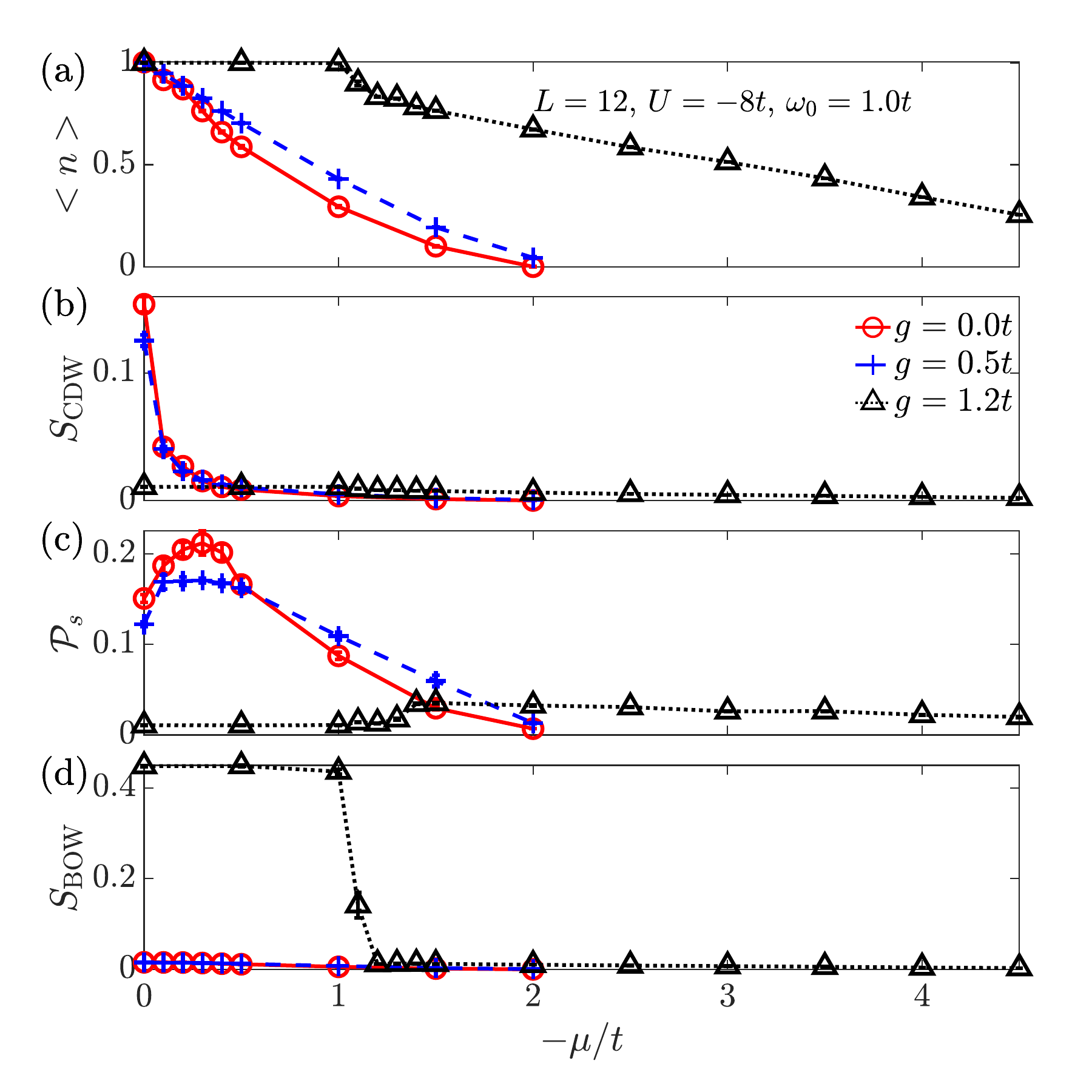} 
    \caption{
        (a) Average density $\langle n \rangle$, (b) CDW structure factor $S_{\mathrm{CDW}}$, (c) $s$-wave pairing correlation $\mathcal{P}_{s}$ and (d) BOW structure factor $S_{\mathrm{BOW}}$ against chemical potential $\mu$.
        For all results, system size $L = 12$, electron-electron coupling $U = -8t$ and phonon frequency $\omega_{0} = 1.0t$.
        At $g = 0.0$, the system is described by the attractive Hubbard model.
    }
    \label{fig:fixU_varyg_b16}
\end{figure}

In the above analysis, we investigated the effects of the electron-electron coupling, $U$, on the SSH BOW. 
Next, we study the effects of the electron-phonon coupling, $g$, on the properties of the attractive Hubbard model.  Specifically,
how $g$ and doping affect the degenerate CDW/$s$-wave pairing phase in the ground state at half-filling, and the superconducting phase (with a finite-$T$ Kosterlitz-Thouless transition~\cite{Kosterlitz1974}).

\begin{figure}[t]
    \centering
    \includegraphics[width=\linewidth]{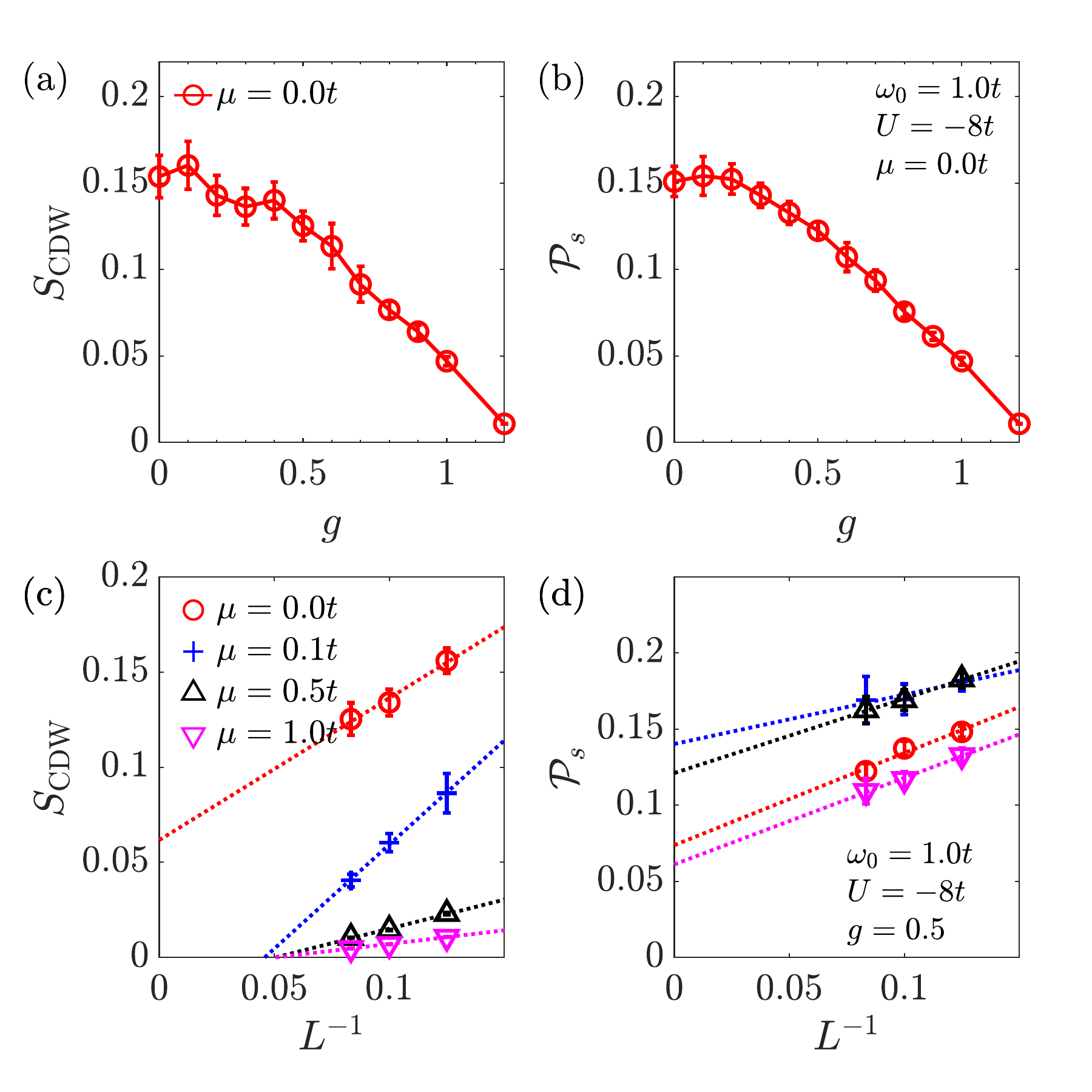} 
    \caption{
        (a) CDW structure factor $S_{\mathrm{CDW}}$ and (b) $s$-wave pairing correlation $\mathcal{P}_{s}$ against electron-phonon coupling $g$ at half-filling.
        (c) Finite-size scaling of $S_{\mathrm{CDW}}$ and (d) $\mathcal{P}_{s}$ at $g=0.5$.
        For all subplots, phonon frequency $\omega_{0} = 1.0\,t$ and electron-electron coupling $U = -8\,t$.  
    }
    \label{fig:fixU_varyg_b16_n1}
\end{figure}

In Fig.~\ref{fig:fixU_varyg_b16}, we plot (a) the average density $\langle n \rangle$, (b) CDW structure factor $S_{\mathrm{CDW}}$, (c) $s$-wave pairing correlation $\mathcal{P}_{s}$ and (d) BOW structure factor $S_{\mathrm{BOW}}$ versus the chemical potential for $L = 12, U = -8t, \omega_{0} = 1.0t$.
At $g = 0.0$, the system is described purely by the attractive Hubbard model.
The degenerate CDW/pairing symmetry is lifted immediately with finite $\mu$:
Away from half-filling, $S_{\mathrm{CDW}}$ falls rapidly (Fig.~\ref{fig:fixU_varyg_b16}(b)) while $\mathcal{P}_{s}$ is maximized at an intermediate density (Fig.~\ref{fig:fixU_varyg_b16}(c)).
At $g = 0.5$, we observe a decrease in both $S_{\mathrm{CDW}}$ (Fig.~\ref{fig:fixU_varyg_b16} (b)) and $\mathcal{P}_{s}$ (Fig.~\ref{fig:fixU_varyg_b16}(c)).
The effect of $g$ on the half-filled attractive Hubbard model is further studied in Fig.~\ref{fig:fixU_varyg_b16_n1}(a, b).
As $g$ increases, both $S_{\mathrm{CDW}}$ and $\mathcal{P}_{s}$, which are known to be equal at half-filling, decrease.
This suggests that the SSH electron-phonon interaction works against the attractive Hubbard interaction at and away from half-filling.
While the attractive Hubbard $U$ and SSH $g$ support different ground state phases, we find no evidence of an intermediate phase for the parameters we have studied.
At $g = 1.2$, the strong SSH electron-phonon interaction dominates and the half-filled ground state becomes the gapped $q = \left( \pi,\pi \right)$ BOW.

In Fig.~\ref{fig:fixU_varyg_b16_n1}(c, d), we show the finite-size scaling of $S_{\mathrm{CDW}}$ and $\mathcal{P}_{s}$ at $g = 0.5$.
Like the pure attractive Hubbard model, $S_{\mathrm{CDW}}$ extrapolates to a finite value only at half-filling $\mu = 0$.
In Fig.~\ref{fig:fixU_varyg_b16_n1}(c), an infinitesimal $|\mu| = 0.1$ is enough to destroy the CDW phase.
In Fig.~\ref{fig:fixU_varyg_b16_n1}(d), $\mathcal{P}_{s}$ extrapolates to a finite value for a large range of $\mu$.
These finite-size scaling results show that the electron-phonon coupling $g$ does not qualitatively change the attractive Hubbard ground state for $ g < g_{c}$. 
At half-filling, the ground state is either the SSH BOW or the degenerate attractive Hubbard CDW/$s$-wave pairing phase.
Away from half-filling, the ground state always exhibits $s$-wave pairing, albeit moderated by the magnitude of the electron-phonon coupling $g$.

\vskip0.05in \noindent 
\underline{{\it Conclusions:}}
In this work, we present a DQMC study of the single-orbital square lattice {\it attractive} SSHH model. 
Because of the lack of sign problem, we can study the interplay between the (SSH) electron-phonon interaction and 
electron-electron interactions {\it in the ground state}. 

At half-filling, we find a first-order quantum phase transition between the BOW phase and the degenerate CDW/$s$-wave pairing phase.
At higher $g$, the ground state remains in the degenerate CDW/$s$-wave pairing phase until the BOW sets in for large enough $g$. 
Away from half-filling, the CDW phase and BOW phase are both unstable.
The ground state is $s$-wave pairing but the order parameter can be reduced by larger electron-phonon interactions. 
In addition, we show the effects of phonon frequency. 
As it decreases, phonon fluctuations increase and the electron-phonon mechanism becomes more robust.
As in the case of the Holstein model, this leads to a reduction in $s$-wave pairing.
Other than pointing out that its magnitude becomes much smaller at lower phonon frequencies, we cannot rule out the possibility of its existence.

The fundamental feature of the SSH Hamiltonian is the modulation of the inter-site hopping by a phonon displacement $X_{ij}$.
A related model in which this electron-phonon interaction and on-site $U$ is present has been suggested as a description of the Kondo effect arising from a vibrating magnetic ion\cite{yotsuhashi2005,hattori2005,mitsumoto2005,hotta2006,hotta2008,fuse2010,oshiba2011,fuse2013}. 
Recent advancements in experimental setups have opened up the possibility to study different variants of the SSH model physically~\cite{MeierGadway2016a, LederWeitz2016, LiuZhang2019a, ZhengPagneux2019a, ChenCheng2022}.
These studies have focused on the topological properties of the static SSH model, which assumes a frozen phonon structure.
Our theoretical work expands further by studying the SSH model with full phonon dynamics.
Moving forward, it will be interesting to investigate further the low $U$ regime, which even in the absence of electron-phonon coupling is much harder to study~\cite{white89a,pruschke2003slater,raczkowski2021local}.
Another area that could benefit from further investigation is the regime of low phonon frequencies which, due to smaller gaps, require larger systems and lower temperatures. 

\vskip0.05in \noindent
\underline{{\it Acknowledgments:}}
We acknowledge fruitful discussions with 
B. Cohen-Stead and S. Johnston. 
D.P. acknowledges support from the National Research Foundation, Singapore, and A*STAR under its Quantum Engineering Programme (NRF2021-QEP2-02-P03).
The work of RTS was supported by the grant DE‐SC0014671 funded by the U.S. Department of Energy, Office of Science. 
The computational work for this Letter was performed on resources of the National Supercomputing Centre, Singapore (NSCC) \cite{nscc}. The Flatiron Institute is a division of the Simons Foundation.

\bibliography{main}

\end{document}